\documentstyle[pra,aps,twocolumn,epsfig]{revtex}

\begin{document}

\title{Time correlations of a laser--induced Bose--Einstein condensate}

\author{Timo Felbinger$^{\dag}$, Luis Santos$^{\ddag}$, 
Martin Wilkens$^{\dag}$ and Maciej Lewenstein$^{\ddag}$}

\address{\dag Universit\"at Potsdam, Fachgruppe Physik, Am Neuen Palais 10, 
D-14469 Potsdam, Germany \\
\ddag Institut f\"ur Theoretische Physik, Universit\"at Hannover,
 Appelstr. 2, D--30167 Hannover, Germany}

\maketitle

\begin{abstract}
We analyze the multi--time correlations of a laser--induced 
Bose--Einstein condensate. We use quantum 
stochastic methods to obtain under certain circumstances 
a Fokker--Planck equation which describes the 
phase--difussion process, and obtain an analytical expression 
of the two--time correlations. We 
perform also quantum Monte Carlo numerical simulations of the correlations, 
which are in good agreement with the 
predicted analytical results.
\end{abstract}
\pacs{32.80Pj, 42.50Vk}

\section{Introduction}

During the last years the combination of laser cooling \cite{Nobel} and evaporative 
cooling methods \cite{BEC} has lead to the achievement of one of the most pursued 
goals since the early days of quantum physics, i.e. the so--called Bose--Einstein 
Condensation (BEC). Such condensation is a direct 
 consequence of the Bose--Einstein statistics, and consists in the fact that the 
ground state of the system becomes macroscopically populated. Although BEC has been 
obtained using the combination of laser and collisional techniques, there are several 
experimental groups \cite{Salomon,Ertmer,Mlynek,Strontium,recent} which currently 
investigate the possibility of obtaining BEC using  all--optical means only. 
In such a case the number of atoms in the trap would not 
decrease during the cooling process, and a non--destructive detection of BEC could be performed by 
simply measuring the fluorescence photons. In addition, laser induced condensation would be easier
to control externally, and could lead to richer effects than collisional processes, as those
employed in evaporative and sympathetic cooling. Finally, the methods employed to 
achieve laser induced 
condensation can be used to design the techniques of pumping the atoms into the condensate 
using spontaneous emission \cite{Spreeuw95,Janicke96,Janicke99,Bar}.

The most sophisticated laser cooling techniques (as Velocity Selective Coherent Population Trapping
\cite{VSCPT} or Raman cooling
\cite{Raman}) are capable to cool atomic samples below the photon--recoil energy,
$E_{R}=\hbar\omega_{R}=\hbar^{2}k_{L}^{2}/2m$, where $k_{L}$ is the laser wavevector, and $m$ is the
atomic mass. Such methods should in principle lead to BEC. However, in using laser cooling to obtain 
the BEC, the reabsorption of spontaneously emitted photons turns to be a 
very important problem. This is because the sub-recoil laser cooling 
techniques are based on dark--state 
mechanisms \cite{Orriols}, and the dark--states are unfortunately not dark respect to the spontaneously 
emitted photons. Therefore,
multiple reabsorptions can increase the system energy by several recoil energies per atom
\cite{Sesko,Dalibard,Burnett,Ellinger}.  It is easy to understand that assuming that the reabsorption cross
section for trapped atoms is the same as in free space, i.e. $\simeq1/k_L^2$, the significance of
reabsorptions increases with the dimensionality \cite{Mlynek,Dalibard}, 
in such a way that the reabsorptions should  not
cause any problem in one dimension. For the case of three--dimensional 
traps several remedies have also been proposed 
\cite{reabs}. In the following we are going to assume that 
our system is one--dimensional. This assumption in the first place has been done for 
simplicity reasons, but at the same time it allows us to avoid the reabsorbtions problem. 
As we discuss in the conclusion section, however, our approach can be used 
in three--dimensional asymmetric traps, for which heating due to reabsorbtions can also be negligible. 

When laser--cooling methods are employed, the thermalization of the system 
occurs with the interaction 
with the laser field and the vacuum modes of the electromagnetic field, and therefore 
atom--atom collisions 
are in principle not necessary. In the following we shall consider an ideal gas 
(see the discussion in Sec.\ 
\ref{sec:Model}). 
In fact, the proper notion of temperature must then  be revised. On the other hand, 
the dynamics of the system becomes more complex, because we have to deal with an open system 
which interacts 
with the laser. In Refs. \cite{Cirac94,Cirac94b}, the quantum dynamics of a laser--cooled ideal gas 
has been studied using second quantization formalism in order to take into account 
the quantum--statistical 
character of the bosons, and also employing quantum stochastical methods 
\cite{Gardinerbook1,Gardinerbook,Carmichaelbook} to take into 
account the coupling with the reservoir provided 
by the vacuum modes of the electromagnetic field.  The equations which describe the dynamics of the 
system have been obtained, and so has the stationary solution, which corresponds for the case of a bosonic system 
with the well--known Bose--Einstein distribution (BED), and under proper conditions the BEC appears.

In order to characterize this laser--induced BEC, the two--time correlations of the condensate amplitude 
turn to be very important quantities, 
because they describe the diffusion of the phase of the condensate. These correlations are very difficult to calculate 
both analytically and numerically, and until now
 no calculation of such correlations has been  published, as far as we know. 
The main aim of the present paper is to present a method to calculate such correlations using 
quasi--probability representations, and in particular $P$--representation \cite{Gardinerbook}. 
We shall also present a 
numerical calculation based on the so--called wave--function Monte Carlo 
method \cite{Dalibard92}, whose results are in good 
agreement with the predicted analytical ones.

At this point it is worth noticing that the model considered in this paper is not 
very realistic since it is one dimensional, and even in the relevant single dimension 
it assumes the, so called, Lamb-Dicke limit, when the size of 
the trap $a$ is smaller than the wavelength of the laser field, $\lambda/2\pi$.
 At the same time the model neglects not 
only elastic collisions, but also all non-elastic losses: two-- and three--body 
non-elastic collisions, as well as photoassociation losses. Realization of such situation 
requires modification of atomic scattering length (as discussed in Section II 
and conclusions)and other precautions, which are possible 
(as discussed in Section II and in more detail in 
conclusions), but not 
easy to achieve experimentally. In the worst case, however, the model is realistic 
for small samples of $N=15-50$ atoms. Cooling of such small sample of atoms to a ground 
state of the trap is of fundamental interest itself, and might be even useful for applications for
 quantum information processing \cite{qip}. We stress, however, that the main achievement 
of this paper concerns  {\it methodological aspects}. Even though the model is somewhat unrealistic, 
it is a paradigm model describing quantum dynamics of the bosonic gas approaching the 
equilibrium. As we mentioned, time correlations are very difficult to calculate 
analytically and numerically in any model of that sort, regardless how realistic it is. 
We develop here the methods to calculate such correlations, 
and these methods are of quite  importance themselves, since they are quite 
universal and can be carried over to more realistic models, 
such as those discussed by us in a series of Refs. \cite{luis}.       

The structure of the paper is as follows. In Sec.\ \ref{sec:Model}, 
we briefly review the calculations already 
presented in Ref.\ \cite{Cirac94}. The $P$--representation formalism is 
introduced in Sec.\ \ref{sec:Prep}. 
The expressions obtained in Sec.\ \ref{sec:Prep}, allows us in Sec.\ \ref{sec:Twot} 
to calculate the two--time correlations 
of the amplitude of the condensate. In Sec.\ \ref{sec:Numerical}, 
we present the employed numerical method, and the comparison between 
numerical and analytical results.
We conclude in Sec.\ \ref{sec:Conclusions} stressing once more methodological 
achievements of this paper, and discussing experimental feasibility of our model.

\section{Model}
\label{sec:Model}
In this section we briefly review the formalism already developed in
Ref.\ \cite{Cirac94}. We consider a system composed of N two--level identical 
bosons in an harmonic trap centered at $\vec R =0$, which are coherently 
driven by a 
standing wave laser field, and also interact with the vacuum modes of the 
electromagnetic field. In the following we use, for simplicity, units with 
$\hbar =1$, and velocity of light $c=1$. As in Ref.\ \cite{Cirac94}, we 
consider four basic approximations: (i) the laser is quasi--resonant 
with a particular transition between two electronic states $|g\rangle$ (ground 
state) and $|e\rangle$ (excited state), and therefore we consider  atoms 
as two--level atoms; (ii) Rotating--wave approximation (RWA) can be used, 
 since the frequency 
of the considered electronic transition ($\omega_0$) and the laser 
frequency $\omega_L$ are considered much larger than any other frequency scale 
of the system; (iii) we treat the atom--field interactions in the dipole 
approximation since the resonant wavelength $\lambda_L \gg a_0$, where $a_0$ is
the typical atomic size; (iv) we neglect the atom--atom collisions, i.e. we 
work in the ideal--gas approximation. In principle, the latter  approximation implies 
that we should deal with a small number of particles ($N<100$), because 
we are going to consider that the trap is in the so--called Lamb--Dicke 
Limit (LDL), where  the, so called,  Lamb--Dicke parameter $\eta=2\pi a/\lambda_L <1$, 
with $a$ being the size of the trap ground state. However, the $s$--wave scattering
length $a_{sc}$ (which governs the atom--atom interactions at low energies) 
can be externally modified, for instance,  using a magnetic field 
\cite{Tiesinga92,Ketterle99},  or a laser field\cite{Fedichev96}, 
and in principle $a_{sc}$ can be made very close to zero \cite{Kettcomment,rub85}, 
allowing the strict validity of the ideal gas approximation.

We limit our discussion to the one--dimensional case. The relevant electronic 
states are then determined by the laser polarization perpendicular to the 
laser wave vector, which in turn can be aligned in the $z$--direction, 
$\vec k _L=(0,0,\omega_L)$. We assume that the atoms can only move in the 
$z$--direction, and are localized in transverse direction, so that practically one can assume 
that they are located at 
$\vec R=(0,0,z)$. The atoms are assumed to occupy the energy eigenstates of the 
harmonic trap potential. We denote this energy levels by $|l,g\rangle$ 
($|m,e\rangle$) for the atoms in the ground (excited) electronic state, 
occupying the level $l$($m$)$=0,1,\dots$ of the harmonic trap.

We introduce also the operators $g_l$, $g_l^{\dag}$ ($e_m$, $e_m^{\dag}$) that 
annihilate or create atoms in the $l$th ($m$th) energy level of the 
ground--state (excited--state) potential. These operators fulfill standard 
commutation relations for bosonic atoms:
\begin{eqnarray}
&&[g_l,g_{l'}^{\dag}]=\delta_{ll'}, \label{bosong} \\
&&[e_m,e_{m'}^{\dag}]=\delta_{mm'}. \label{bosone}
\end{eqnarray}

Using standard quantum stochastic methods 
\cite{Gardinerbook1,Gardinerbook,Carmichaelbook}
one can eliminate the vacuum modes of the electromagnetic 
field, 
and after 
performing a systematic expansion in the Lamb--Dicke parameter $\eta$ one obtains \cite{Cirac94} the master equation (ME), 
in the frame rotating with the laser, valid up to the order ${\cal O}(\eta^2)$:
\begin{equation}
\dot\rho=-i[{\cal H}_a+{\cal H}_{las},\rho]+{\cal L}\rho,
\label{me1}
\end{equation}
where the atomic part of the Hamiltonian takes the form
\begin{equation}
{\cal H}_a=\sum_{m=0}^{\infty}\omega_m g_m^{\dag}g_m+
\sum_{m=0}^{\infty}(\omega_m-\delta) e_m^{\dag}e_m,
\label{Ha}
\end{equation}
with $\delta=\omega_L-\omega_0$ being  the laser detuning. In Eq.\ (\ref{Ha}), 
the center--of--mass potentials for ground and excited atoms can be well 
approximated by harmonic potentials of frequency $\omega_g=\omega_e=\omega$ 
plus a small anharmonicity so that the energy levels are $\omega m + 
\omega\alpha m^2$ with $m=0,1,\dots$ and $\alpha\ll 1$. From now on we use the 
same indices for the energy levels in the ground-- and excited potentials.

The Hamiltonian of the atom--laser interaction takes the form:
\begin{eqnarray}
{\cal H}_{las}&=&\eta\frac{\Omega}{2}\sum_m^{\infty}\sqrt{m+1}
\left ( e_{m+1}^{\dag}g_m+ e_{m}^{\dag}g_{m+1}+H.c.\right ) \nonumber \\
&+&{\cal O} (\eta^3),\label{Hlas}
\end{eqnarray}
where $\Omega$ denotes the Rabi frequency describing the laser--atom interaction.

The spontaneous emission part in Eq.\ (\ref{me1}) takes the form:
\begin{eqnarray}
{\cal L}\rho&=&\frac{\Gamma}{2}\sum_{m,m'}
\left ( 2 g_m^{\dag}e_m\rho e_{m'}^{\dag}g_{m'}-
e_{m'}^{\dag}g_{m'}g_m^{\dag}e_m\rho \right \delimiter 0 \nonumber \\
&-&\left\delimiter 0 \rho e_{m'}^{\dag}g_{m'}g_m^{\dag}e_m \right ) + 
{\cal O}(\eta^2),
\label{L}
\end{eqnarray}
where $\Gamma$ is the natural line-width of the considered transition.

If we locate the trap at the node of the laser standing wave, the excitation 
provided by the laser then become weak, provided that $\Omega$ is moderate. 
We expect then that at a given instant no more than one atom will be excited. 
The system is then characterized by two distinct time scales: the fast one, 
determined by $N\Gamma$, $\omega$ and $\delta$; and the slow one, 
characterized by $\eta^2\Omega^2 /N\Gamma,\eta^2\Omega^2 /(\omega\pm\delta)$, 
which describes the jumps between the various ground--state levels. In such a 
case one can use 
standard adiabatic elimination techniques \cite{Gardinerbook1} to remove 
excited state populations. Physically, this reflects the fact that a 
combined process of excitation and (relatively rapid) spontaneous decay causes a
redistribution of atomic population among different levels of the 
ground--state trap. The ME after the adiabatic elimination of the excited 
states becomes:
\begin{equation}
\dot\rho=\sum_m^{\infty}{\cal L}_m\rho,
\label{me2} 
\end{equation}
where the coupling between the levels $m$ and $m+1$ \cite{LDLcomment} is 
given by \cite{Cirac94}:
\begin{eqnarray}
{\cal L}_m\rho&=&\frac{\Gamma_-}{2}(m+1)
(2A_m\rho A_m^{\dag}-A_m^{\dag}A_m\rho-\rho A_m^{\dag}A_m) \nonumber \\
&+& \frac{\Gamma_+}{2}(m+1)
(2A_m^{\dag}\rho A_m-A_m A_m^{\dag}\rho-\rho A_m A_m^{\dag}),
\label{Lm}
\end{eqnarray}
with $A_m=g_m^{\dag}g_{m+1}$, and 
\begin{equation}
\Gamma_{\pm}=\Gamma (\eta\Omega /2)^2
\frac{1}{(N\Gamma /2)^2+(\omega\mp\delta)^2}.
\label{Gam}
\end{equation}

It is easy to check \cite{Cirac94} that for $\delta<0$ there exists an exact steady state 
solution of ME (\ref{me2}) which has the canonical form:
\begin{equation}
\rho_{st}=\frac{1}{Z}q^{\sum_m m g_m^{\dag}g_m},
\label{stat}
\end{equation}
where $q=\exp(-\omega/k_B T)=\Gamma_+ / \Gamma -$, and $Z$ is the canonical 
partition function. For the particular case of a one--dimensional harmonic 
potential the partition function can be derived in closed form \cite{Gajda97}:
\begin{equation}
Z=\prod_{j=1}^N\frac{1}{1-q^j}.
\label{Z}
\end{equation}
The thermodynamics of the system is determined by the Helmholtz free energy 
$F=-k_B T \ln (Z)$. The canonical version of the chemical potential is then given 
by:
\begin{equation}
\mu\equiv(\partial F/\partial N)_{T,\omega}=k_B T \ln[1-q^N ].
\label{mu}
\end{equation}
There is a non--zero temperature \cite{Wilkens97}
\begin{equation}
T_c=\frac{\hbar\omega}{k_B}\frac{N}{\ln (N)},
\label{Tc}
\end{equation}
below which $\mu$ is effectively zero.
$q$ can be conveniently re--written in terms of this critical temperature:
\begin{equation}
q=\exp \left (- \frac{\ln (N)}{N}\frac{T_c}{T} \right ).
\label{q}
\end{equation}

\section{P--Representation}
\label{sec:Prep}

In the following section, we are going to employ ME (\ref{me2}) to calculate 
the two--time correlations of the ground state in the stationary regime, i.e. correlations of the
form $\langle g_0^{\dag}(t) g_0(0) \rangle $. In order to do that
it is convenient  to use the so--called 
Glauber--Sudarshan $P$--representation \cite{Gardinerbook}, which is defined as:
\begin{eqnarray}
\rho(t)&=&\int d^2z_0d^2z_1\dots P(z_0,z_0^{\ast};z_1,z_1^{\ast};\dots) \nonumber \\
&\times&|z_0,z_1,\dots\rangle\langle z_0,z_1,\dots |. \label{Prep}
\end{eqnarray}
where $|z_0,z_1,\dots\rangle$ are coherent states. 
After transforming into $P$--representation, Eq.\ (\ref{me2}) becomes:
\begin{equation}
\dot P=\sum_{m=0}^{\infty}\hat S_m P,
\label{Pdot}
\end{equation}
where the transition between $m$ and $m+1$ is given by the operator:
\begin{eqnarray}
&&\hat S_m=\frac{\Gamma_-}{2}(m+1) \nonumber \\ 
&& \left \{ 2\left [ |z_m|^2-
\left ( \frac{\partial}{\partial z_m}z_m+\frac{\partial}{\partial z_m^{\ast}}z_m^{\ast} \right )+
1 +\frac{\partial^2}{\partial z_m \partial z_m^{\ast}} \right ] |z_{m+1}|^2  \right\delimiter 0\nonumber \\
&&-\left ( |z_{m+1}|^2-\frac{\partial}{\partial z_{m+1}}z_{m+1} \right )
 \left ( |z_{m}|^2-\frac{\partial}{\partial z_{m}}z_{m} +1 \right ) \nonumber \\
&&-\left ( |z_{m+1}|^2-\frac{\partial}{\partial z_{m+1}^{\ast}}z_{m+1}^{\ast} \right )
 \left ( |z_{m}|^2-\frac{\partial}{\partial z_{m}^{\ast}}z_{m}^{\ast} +1 \right ) \nonumber \\
&& 2q\left [ |z_{m+1}|^2-
\left ( \frac{\partial}{\partial z_{m+1}}z_{m+1}+\frac{\partial}{\partial z_{m+1}^{\ast}}z_{m+1}^{\ast} \right )+
1 + \right \delimiter 0 \nonumber \\
&&\left\delimiter 0\frac{\partial^2}{\partial z_{m+1} \partial z_{m+1}^{\ast}} \right ] |z_m|^2  \nonumber \\
&&-\left ( |z_m|^2-\frac{\partial}{\partial z_m}z_m \right )
 \left ( |z_{m+1}|^2-\frac{\partial}{\partial z_{m+1}}z_{m+1} +1 \right ) \nonumber \\
&&-\left \delimiter 0 \left ( |z_m|^2-\frac{\partial}{\partial z_m^{\ast}}z_m^{\ast} \right )
 \left ( |z_{m+1}|^2-\frac{\partial}{\partial z_{m+1}^{\ast}}z_{m+1}^{\ast} +1 \right ) 
\right \}
\label{Sm}
\end{eqnarray}

Let us define the reduced $P$--representation for the ground state of the trap as:
\begin{equation}
P_0(z_0,z_0^{\ast};t)=\int d^2 z_1 d^2 z_2 \dots P(z0,z0^{\ast};z1,z1^{\ast};\dots;t).
\label{P0}
\end{equation}
The equation which describes the dynamics of $P_0$, takes the form:
\begin{eqnarray}
\dot P_0&=&\frac{\Gamma_-}{2}q 
\left ( \frac{\partial}{\partial x_0}x_0+\frac{\partial}{\partial x_0^{\ast}}x_0^{\ast}\right ) P_0 
\nonumber \\
&+& \frac{\Gamma_-}{2}(q-1) 
\left ( \frac{\partial}{\partial x_0}x_0+\frac{\partial}{\partial x_0^{\ast}}x_0^{\ast}\right ) P_1
\nonumber \\
&+&\frac{\Gamma_-}{N}\frac{\partial^2}{\partial x_0 \partial x_0^{\ast}}P_1.
\label{P0dot} 
\end{eqnarray}
where $P_1=\int d^2 z_1 d^2 z_2 \dots |z_1 |^2 P$. We have also introduced here the convenient 
notation  
$z_0=\sqrt{N}x_0$.

One can prove (see Appendix \ref{app:apenA}) that for times $t>[\Gamma_- (1-q) N|x_0|^2]^{-1}$, one can 
adiabatically eliminate the excited trap states, and retrieve a closed Fokker--Planck equation (FPE) for the 
reduced $P_0$:
\begin{eqnarray}
&&\dot P_0 = \frac{\Gamma_- q}{N(1-q)}\left \{ \frac{\partial ^2}{\partial x_0\partial x_0^{\ast}}
\right\delimiter 0 \nonumber \\  
&&\left\delimiter 0
-\frac{1}{4}\left [ \frac{\partial}{\partial x_0}x_0+\frac{\partial}{\partial x_0^{\ast}}x_0^{\ast}\right ]
\left [ \frac{\partial}{\partial x_0}x_0+\frac{\partial}{\partial x_0^{\ast}}x_0^{\ast}\right ]\frac{1}{|x_0|^2}
\right \} P_0.
\label{P0dot2}
\end{eqnarray}
Changing into the more convenient variables $z_0=r\exp(i\phi)$, the FPE (\ref{P0dot2}) becomes:
\begin{equation}
\dot P_0=\frac{{\cal K}}{r^2}\frac{\partial^2}{\partial\phi ^2}P_0,
\label{P0dot3}
\end{equation}
with
\begin{equation}
{\cal K}=\frac{\Gamma_- q}{4(1-q)}.
\label{K}
\end{equation}

Eq. (\ref{P0dot3}) is one of the central results of this paper, and has a simple physical 
interpretation as an equation describing diffusion of the phase of the condensate wave function.
Interestingly, even though we are dealing here with an ideal gas approaching the thermal 
equilibrium, the equilibrium state has temporal properties characteristic for the states 
obtained via spontaneous breaking of the $U(1)$ phase symmetry, such as in the theory of laser, 
or in general any theory of second order phase transitions in which the effective 
potential for the order parameter below the transition 
temperature has a Mexican hat shape \cite{WilkensMex}. In particular, the phase diffusion rate
is here inverse proportional to $r^2$, i.e. to the number of atoms in the condensate.      

\section{Two--time correlations of the condensate}
\label{sec:Twot}

The solution of the FPE (\ref{P0dot3}) can be obtained using the following 
Green function:
\begin{eqnarray}
&&G(r,\phi,t || r_o,\phi_0, t=0)= \nonumber \\
&&\frac{1}{\pi r_0}\delta(r-r_0)[\frac{1}{2}+\sum_{n=1}\cos n(\phi-\phi_0)
e^{-n^2\frac{{\cal K}}{r_0^2}t}] .
\label{Green}
\end{eqnarray}
Using this  we can already calculate the time correlation:
\begin{eqnarray}
  &&\langle g_0^{\dag l}(t)g_0^l(0) \rangle \nonumber \\ 
  &&= \int_0^{\infty}\int_0^{2\pi}rdrd\phi\int_0^{\infty}\int_0^{2\pi}r_0dr_0d\phi_0 \nonumber \\
  &&\times G(r,\phi,t || r_o,\phi_0, t=0)P_0(r_0,\phi_0)r^lr_0^l e^{-il(\phi-\phi_0)}\nonumber \\
  &&=\int_0^{\infty}\int_0^{2\pi}rdrd\phi r^{2l}e^{-l^2\frac{{\cal K}}{r^2}t}P_0 (r,\phi) \nonumber \\
  &&=\langle : (g_0^{\dag}g_0)^l e^{-l^2 {\cal K}t/g_0^{\dag}g_0} : \rangle,
  \label{g1}
\end{eqnarray}
where ``:'' denotes normal ordering.
In the following we just consider the case of an exponent $l=1$.
We expand the exponent, assuming that for temperatures $T$ sufficiently below $T_c$, fluctuations of
$n_0 = g_0^\dag g_0$ are small:
\begin{eqnarray}
    {{\cal K} t \over n_0 }
  \,&=&\,
    {{\cal K} t \over \langle n_0\rangle - ( \langle n_0\rangle - n_0 ) } \nonumber \\
  \,&\approx&\,
    {{\cal K} t \over \langle n_0 \rangle } \left(
        1 + \left( 1 - { n_0 \over \langle n_0\rangle } \right)
      \right) \nonumber \\
  \,&=&\,
    { 2 {\cal K} t \over \langle n_0 \rangle } 
      - {{\cal K} t \over \langle n_0 \rangle^2 } n_0
      \,.
\end{eqnarray}
Using this expression and the identity
  $:\exp( \xi g_0^\dag g_0 ): \ = \ \exp( \ln(\xi+1) g_0^\dag g_0 )$,
Eq.\ (\ref{g1}) can be rewritten in the form:
\begin{eqnarray}
  && \!\!\! \langle g_0^{\dag}(t)g_0(0) \rangle \nonumber \\ 
  && = \langle : g_0^{\dag}g_0 \exp \left[
         - \frac{2{\cal K}t}{\langle n_0\rangle}
         + \frac{{\cal K}t}{\langle n_0\rangle^2} g_0^{\dag}g_0
       \right] : \rangle \nonumber \\ 
  && = e^{-2{\cal K}t/\langle n_0 \rangle   } \langle
         n_0 \exp\left[
            (n_0-1) \ln( 1 + {\cal K} t / \langle n_0 \rangle^2 )
         \right]
       \rangle
       \,.
  \label{g1_3}
\end{eqnarray}
Note that from Eq.\ (\ref{g1_3}), 
it is clear that the correlations depend on the fluctuations of 
the condensate fraction, which are not correctly described using the 
usual text--book treatment of the system based on a grand--canonical 
ensemble \cite{Gajda97,Wilkens97,Politzer96,Grossmann97,Navez97}.
In order to calculate 
the correlations,  it is thus necessary to calculate the required averages 
using physically sound ensemble, which in this case is 
 the canonical one. Using canonical ensemble we have to determine the 
 probabilities  to have $n_0$ 
particles in the ground state, given a 
total number $N$ of particles \cite{Wilkens97}:
\begin{equation}
P_0^{CN}(n_0 |N)=q^{N-n_0}\prod_{j=N-n_0+1}^N (1-q^j).
\label{Pcn}
\end{equation}
Therefore:
\begin{eqnarray}
  && \!\!\! \langle(g_0^{\dag}(t))(g_0(0)) \rangle \nonumber \\
  && = e^{-2{\cal K}t/\langle n_0 \rangle}
         \sum_{n_0=0}^{N}P_0^{CN}(n_0 |N) n_0 \left(
             1+\frac{{\cal K}t}{\langle n_0 \rangle^2}
          \right)^{n_0-1}
  \,.
  \label{g1_4}
\end{eqnarray}
The above closed analytic formula is another central result of this paper. 

\section{Numerical results}
\label{sec:Numerical}

Numerical simulations of the system under consideration 
can be performed using Quantum Monte Carlo techniques. In our case, it is 
a quite difficult  task due to the large size of the Hilbert space which has to be simulated, 
even if only a relatively small total particle number $N$ and a low cutoff of 
the trap level structure is chosen.

To avoid having to compute the dynamics of the huge density matrix of 
this system, we applied a stochastic wave function method, 
which replaces the deterministic evolution of the density matrix 
(following Eq.(\ref{me2})) by a stochastic evolution of an ensemble
of state vectors \cite{wwwaddress}.

A naive application of this algorithm to simulate (\ref{me2}) would lead to a stochastic
jump process in which the operators $A_m$ and $A_m^\dag$ act as jump operators.
This process would represent the density matrix as an
ensemble of simultaneous eigenvectors of all occupation
number operators $n_m = g_m^{\dag} g_m$, causing large fluctuations of the phase which
we are interested in. Consequently, the Monte Carlo simulations would require a very
large number of realizations.

We found that convergence can be improved by rewriting the Liouvillian operators
given in Eq.~(\ref{Lm}) in the following way:
  \begin{eqnarray}
    \, && \! {{\cal L}}_m \rho
        \nonumber \\
    &&=
        \frac{\Gamma_-}{2}(m\mathord+1)
          (2{A_m^{(+)}}\rho {A_m^{(+)}}^{\dag} - {A_m^{(+)}}^{\dag}{A_m^{(+)}}\rho - \rho {A_m^{(+)}}^{\dag}{A_m^{(+)}} )
        \nonumber \\
    &&+ \frac{\Gamma_-}{2}(m\mathord+1)
          (2{A_m^{(-)}}\rho {A_m^{(-)}}^{\dag} - {A_m^{(-)}}^{\dag}{A_m^{(-)}}\rho - \rho {A_m^{(-)}}^{\dag}{A_m^{(-)}} )
        \nonumber \\
    &&+ \frac{\Gamma_+}{2}(m\mathord+1)
          (2{A_m^{(+)}}^{\dag}\rho {A_m^{(+)}} - {A_m^{(+)}}{A_m^{(+)}}^{\dag}\rho - \rho {A_m^{(+)}} {A_m^{(+)}}^{\dag} )
        \nonumber \\
    &&+ \frac{\Gamma_+}{2}(m\mathord+1)
          (2{A_m^{(-)}}^{\dag}\rho {A_m^{(-)}} - {A_m^{(-)}}{A_m^{(-)}}^{\dag}\rho - \rho {A_m^{(-)}} {A_m^{(-)}}^{\dag} )
        \,,
    \label{LmAlternative}
  \end{eqnarray}
  with operators
  $A_m^{(\pm)} := {1\over \sqrt{2}} ( A_m \pm \eta_m)$,
  where the $\eta_m$ are c-numbers which are chosen to be of the same order of magnitude as the 
 mean occupation number $\sqrt{ \langle n_m n_{m+1} \rangle }$, so that $\eta_m\approx A_m$.
 Eq. (\ref{LmAlternative}) has an advantage that while it 
is, of course, reproducing the same master equation Eq.~(\ref{me2}), it 
  leads to a different stochastic jump process, in which the $A_m^{(\pm)}$ act as jump operators.
  This process represents $\rho$ as an ensemble of vectors which are, in general, superpositions of
  different occupation number eigenstates, and fluctuations of $g(t)$ within this ensemble
  are much smaller. This observation is yet another important result of this paper.

  Due to computational constraints we have simulated the case of $N=10$ particles, 
confined to the 5 lowest levels of the harmonic trap, for different
 values of the temperature parameter $q=\Gamma_+ / \Gamma_-$.
  For each value of $q$, the function $g(t)$ was estimated by averaging over 1500 trajectories.

\begin{figure}[ht]
\begin{center}
\epsfxsize=5.0cm
\hspace{0mm}
\psfig{file=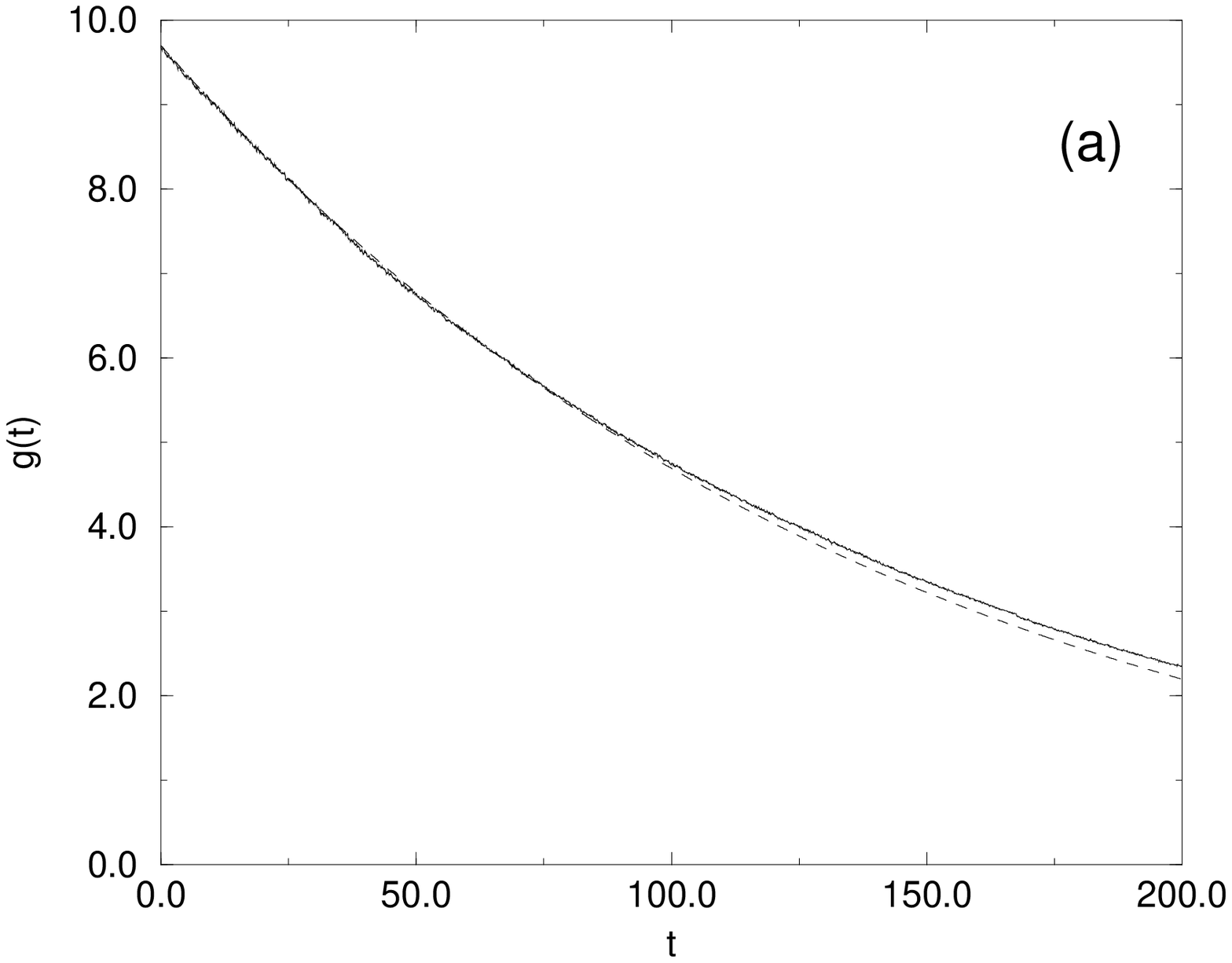,width=5.0cm}
\psfig{file=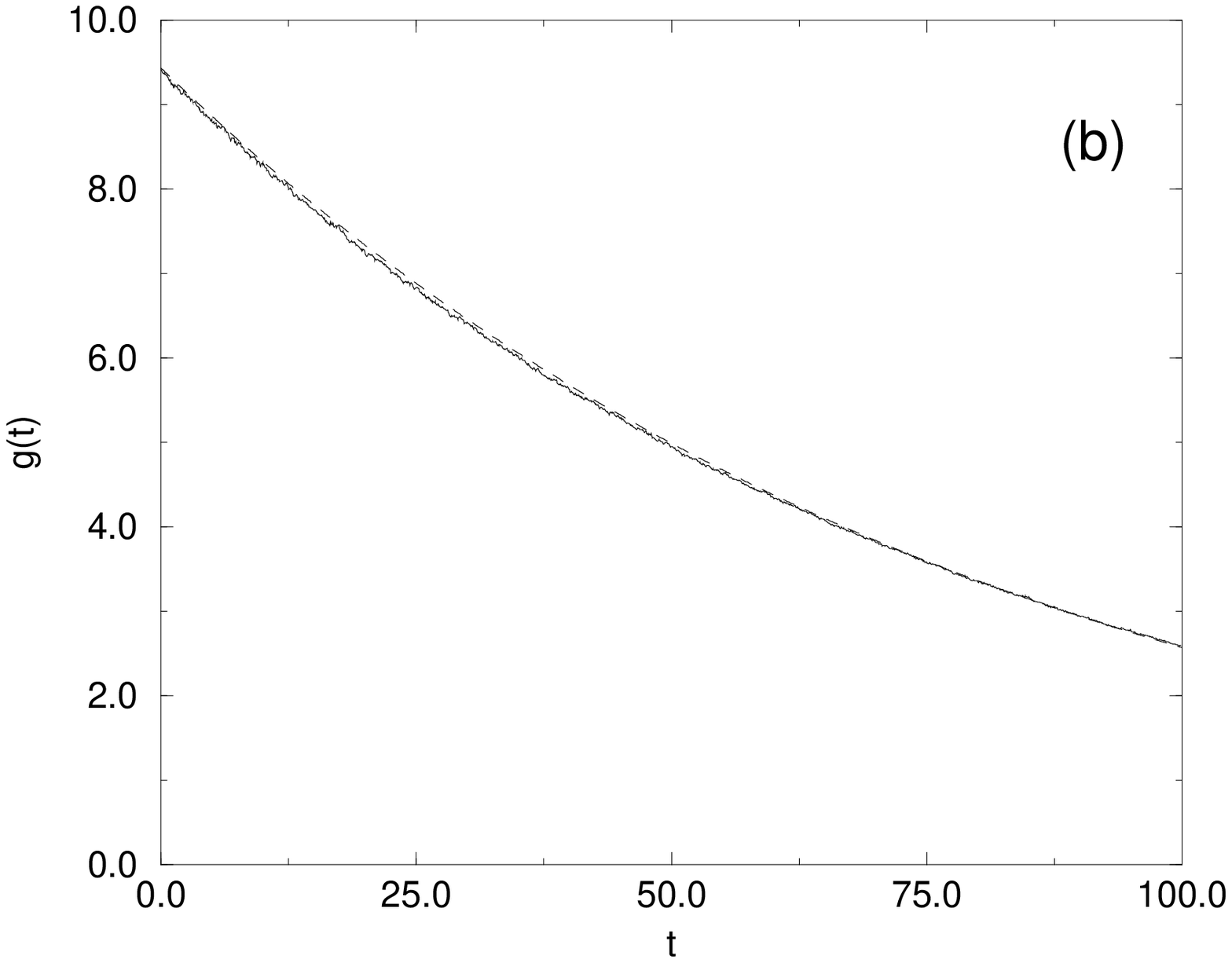,width=5.0cm}\\[0.1cm]
\psfig{file=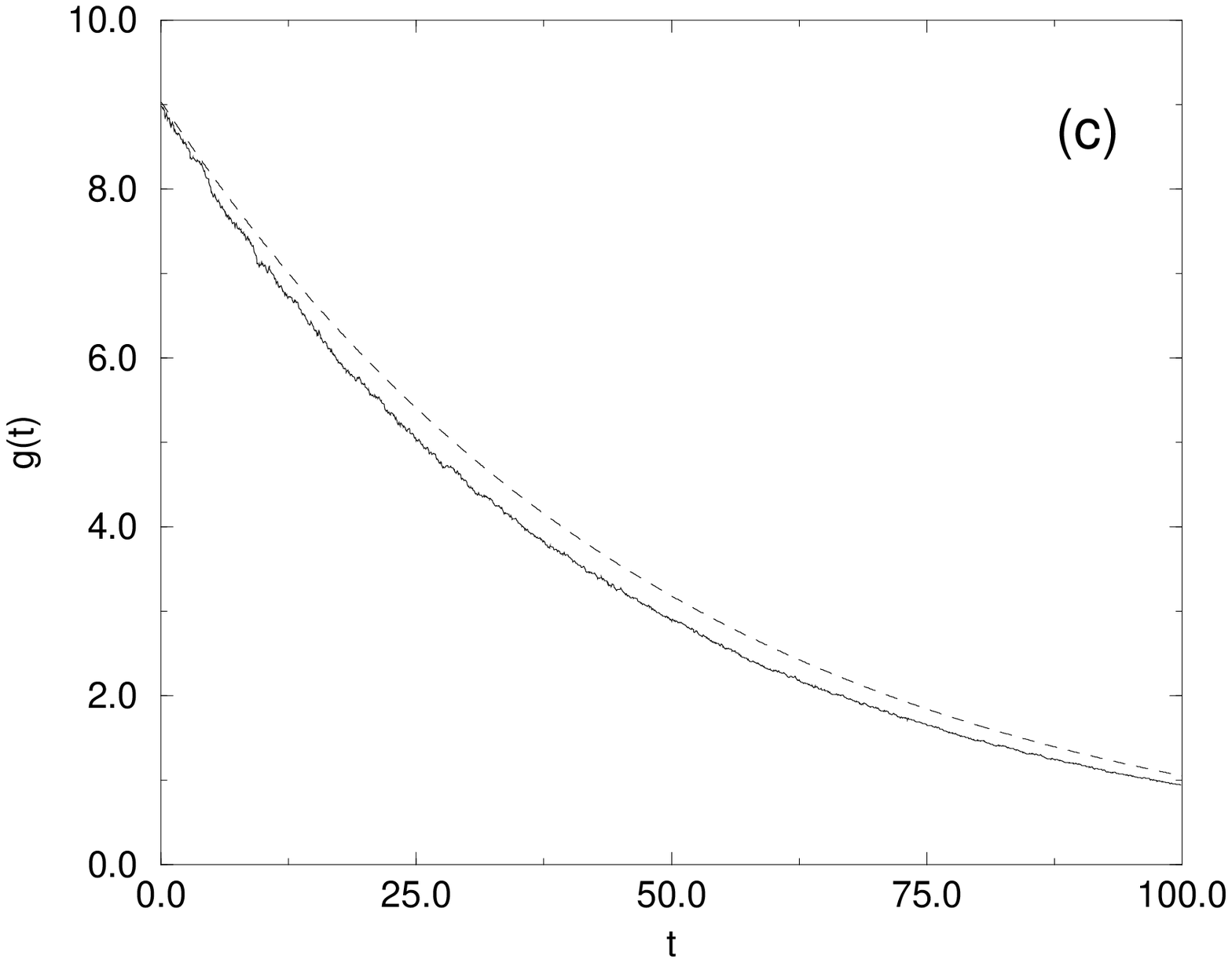,width=5.0cm}
\psfig{file=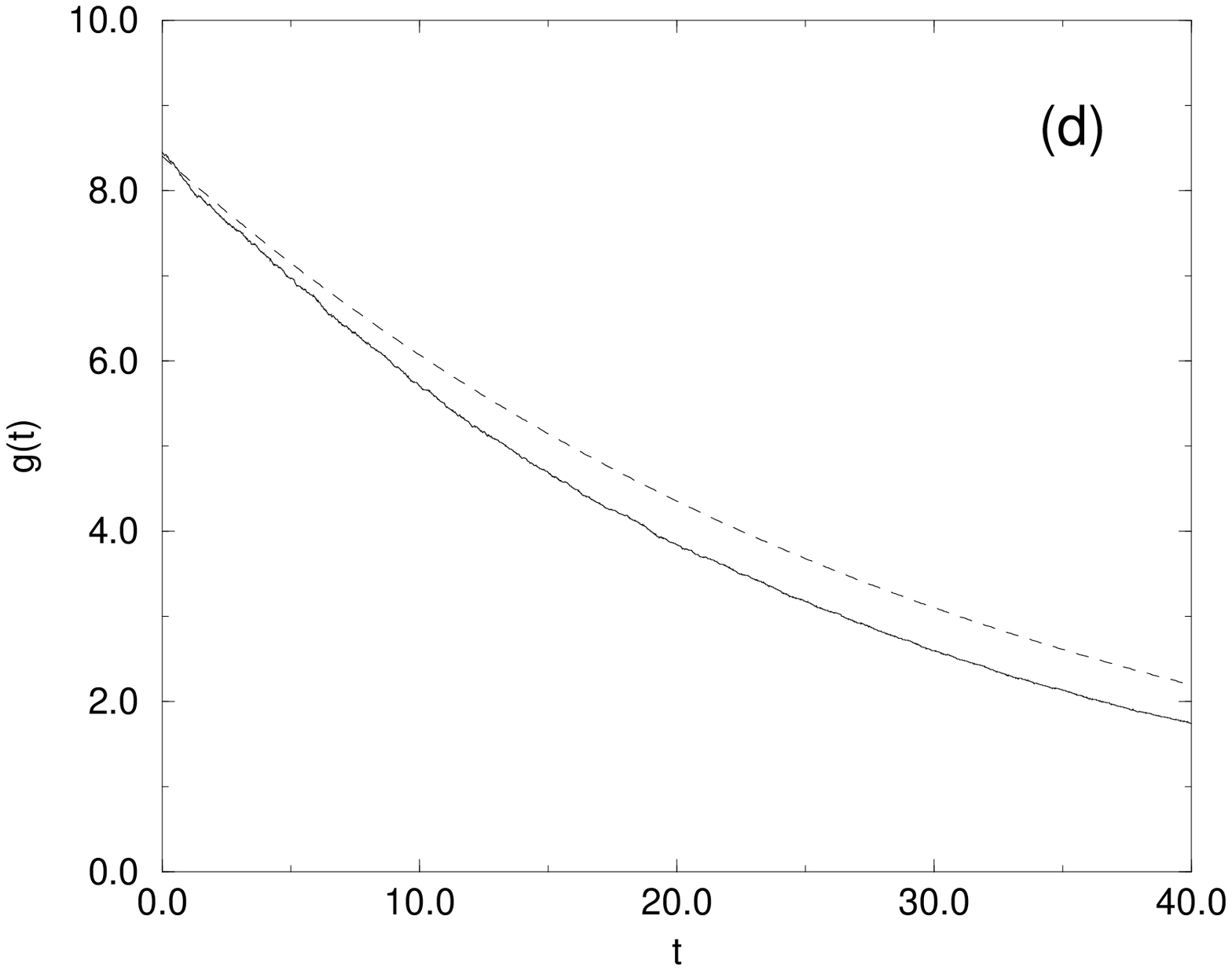,width=5.0cm}\\[0.1cm]
\caption{Comparison between the analytic approximation
described by  
Eq. (\ref{g1_4}) (dashed line),
 and the results of the numerical simulations (solid line), for different 
values of  $q=\Gamma_+ / \Gamma_-$. In the simulations we have chosen the total 
particle number $N=10$. The time scale is in units of $\Gamma_-^{-1}$. 
The numerical results have been obtained by averaging over 1500
trajectories for each value of $q$.
}
\label{plots}
\end{center}
\end{figure}

Figure (\ref{plots}) shows the results of the numerical simulation, compared to the
analytic formula (\ref{g1_4}), for different values of $q$. Quite remarkably, the 
analytical result (based in the approximation of 
$\langle n_0\rangle\gg N-\langle n_0\rangle$) is in very good agreement with 
the numerical results even for $N$ equal just 10 particles (for larger number $N$ the 
agreement should be even better). As  expected, 
the agreement is much better for low temperatures than for high temperatures, 
where the numerical simulation indicates a faster decay of $g(t)$ 
than that predicted by the low-temperature approximation (\ref{g1_4}).

\section{Conclusions}
\label{sec:Conclusions}

In this paper we have studied quantum dynamics of a Bose gas in a trap undergoing sideband
laser cooling in the Lamb-Dicke limit. The master equation describing the dynamics 
of the system can be regarded as a paradigm equation for collective cooling dynamics. 
One of the difficult problems associated with such dynamics consists in calculating 
time dependent correlation functions, such as for instance 
those that describe temporal phase fluctuations of the Bose condensate. In this paper 
we have presented a solution to this problem. Our main results should be regarded 
from the methodological point of view:
\begin{itemize}
\item We have formulated an analytic method to describe temporal correlations based on 
an expansion of the master equation valid at low temperatures, 
when a large number $N_0$ of  particles are in the condensate. The expansion parameter 
in our approach is $1/N_0$. The expansion can be, and already has been applied for other
models of the Bose gas dynamics, that describe more realistic physical situations 
(see Refs. \cite{luis}).

\item We have formulated a numerical method to calculate the time correlations which modifies the 
jump processes involved in the master equation in such a way that the 
corresponding Quantum Monte Carlo simulations are much more stable and require averaging over
much less number of quantum trajectories to achieve good accuracy. The method proposed 
is general, and can be applied not only for the present model, but also for 
more realistic related models. 

\item Analytic and numerical results agree very well even 
at the border of the validity of the analytic theory ($N_0\simeq 10$).

\item Collective laser cooling leads at low temperature to phase diffusion in 
a Mexican hat potential. This result is also general, 
and holds for other more realistic models of the collective laser cooling.

\end{itemize}

The above listed methodological results provide the main value of the present paper. It is,
nevertheless, interesting to speculate whether the consider model is purely academic, or 
whether it can be realized experimentally. We shall argue now that 
it can be realized for small number ($N=10-50)$) of particles.  

Let us discuss step by step the most relevant approximations used in this paper.
\begin{itemize}

\item {\it 1 Dimension.} This approximation was done mainly for technical reasons. The model can be
easily generalized to describe condensation in a three--dimensional trap. In that case, additional 
precautions should be taken into account to avoid the reabsorbtion problem. In particular, 
if the width of the excited states is smaller than $\hbar\omega$, we were thus working not 
only in the LDL limit, but also in the {\it festina lente} regime \cite{reabs}. Extending the dynamics 
to three dimensions in an asymmetric trap will in this regime not introduce any kind of 
dangerous reabsorbtion problems. In fact, the dynamics will consist of 3 independent dynamics 
corresponding to the cooling in the $x$, $y$, and $z$ direction. 

\item{\it Absence of elastic collisions}. As we have mentioned, this requires that the atomic 
density should be sufficiently small, or alternatively that the scattering length should be modified 
to very low values. The first possibility is not interesting, because we require also that the LDL
conditions are fulfilled, which means that $a$ should be of the order of at most $0.1\mu$m.
The condition to be fulfilled is $4\pi N\hbar^2 a_{sc}/mV< \hbar\omega$, with the 
effective condensate volume $V=(2\pi)^{3/2}a^3$.
For the parameter 
considered this gives gives $N\zeta<20$, where $\zeta$ is the modification
factor of the scattering length.

\item{\it Absence of two body inelastic collisions} This problem has a simple remedy. The cooling 
and condensation should take place in a dipole trap, 
and occur in the {\it electronic ground state} of the atoms. Two body 
inelastic processes are then completely suppressed.

\item{\it Absence of three body inelastic processes} Three body losses can be typically neglected
provided the density is less than $10^{15}$/cm$^3$. For $a=0.1\mu$m that requires
$N<(2\pi)^{3/2}\simeq 15$. If the three body losses are modified in a corresponding way 
as elastic  collisions (which seems to be the case for $^{85}$Rb \cite{rub85}), then the corresponding 
condition is much less restrictive $N\zeta^3<15$.

\item{\it Absence of photoassociation losses} This can be reduced  by using red detuned laser
tuned in between the molecular resonances \cite{Fedichev96}. Even is such a case, 
photoassociation losses become relevant when the density reaches the limit $10^{15}$/cm$^3$,
i.e. in our case for $N<(2\pi)^{3/2}\simeq 15$. It seems likely, that this estimate can be 
improved significantly when the laser is tuned below the minimum of the molecular 
transition. While it can hardly be though of for the direct transition, we should stress that 
laser transition considered  in this paper can be equally well regarded as a Raman transition. 
In such case, tuning of the stimulated two-photon transition  below the minimum of the 
molecular resonance is possible. 

\end{itemize}

Summarizing, we see that even in the worst case (no modifications of the scattering length, 
no special precautions regarding photoassociation losses) our model should be valid for $N\simeq 10-15$
particles. Additional precautions can allow to extent the validity of the model to significantly 
larger values of $N$.

\section{Acknowledgments}

We thank P. Navez for discussions.
We acknowledge  support from Deutsche Forschungsgemeinschaft (SFB
407) and the EU through the TMR network ERBXTCT96-0002.

\appendix
\section{Adiabatic elimination of the non--condensed states}
\label{app:apenA}
In this Appendix we present in detail the calculations which allow us to transform 
Eq.\  (\ref{P0dot}) into Eq.\ (\ref{P0dot2}). First, we show that our problem can 
be reduced to a two--level system formed by the level $0$ and $1$ of the trap. Then, we 
adiabatically eliminate the level $1$.

\subsection{Two--level system}

First, we shall analyze the dynamics of $P_1$. From 
(\ref{Pdot}) and (\ref{Sm}) one obtains that
\begin{eqnarray}
&&\dot P_1=\frac{\Gamma_-}{2}\left [\frac{2}{N}\nabla_0^2+(q-1)\hat l_0\right ]P_1^2 \nonumber \\
&& +\frac{\Gamma_-}{2}\left [ 2(q-1)N|x_0|^2+(2q+1)\hat l_0 -2\right ]P_1 \nonumber \\
&& +\Gamma_- qN|x_0|^2 P_0 \nonumber \\
&& +2\Gamma_- \left [ (1-q) P_{1,2}^{1,1}-qP_1+P_2\right ],
\label{P1dot}
\end{eqnarray}
where we have used the notation:
\begin{eqnarray}
&&\nabla_0^2=\frac{\partial^2}{\partial x_0 \partial x_0^{\ast}}, \\
&&\hat l_0 = \frac{\partial}{\partial x_0}x_0+\frac{\partial}{\partial x_0^{\ast}}x_0^{\ast}, \\
&&P_{1,2}^{j,k}= \int d^2 z_1 d^2 z_2 \dots |z_1|^{2j} |z_2|^{2k} P, 
\end{eqnarray}
and $P_1^2=P_{1,2}^{2,0}$, $P_2=P_{1,2}^{0,1}$.
Let us analyze in detail the last line in the RHS of Eq. (\ref{P1dot}), which comes from the contributions 
of $\hat S_1$ in Eq.\ (\ref{Pdot}), i.e. the contributions given by the transitions $1\leftrightarrow 2$. For 
a temperature sufficiently below $T_c$, and sufficiently large N, the atoms in the non--condensed states of the trap 
form a so--called Maxwell--Demon (MD) ensemble \cite{Navez97}, i.e. an ensemble which exchanges particles with 
a reservoir provided by the condensate without exchanging the energy. 
Therefore, the excited states can be considered as: (i) independent of the population of 
the ground state; (ii) decorrelated among each other. Due to these properties:
\begin{equation}
P_{1,2}^{j,k}\simeq  \langle n_1^j \rangle_{GC}  \langle n_2^k \rangle_{GC},
\end{equation}
where  the subindex $GC$ means that the averages are calculated in the grand canonical ensemble. 
It is well known that these averages have the simple form:
\begin{equation}
\langle n_j\rangle_{GC}=\frac{q^j}{1-q^j}
\end{equation}
Therefore, the last line in 
the RHS side of Eq.\ (\ref{P1dot}) becomes:
\begin{equation}
2 \Gamma_- \left [(1-q)\langle n_1\rangle_{GC}\langle n_2\rangle_{GC}-q\langle n_1\rangle_{GC}+\langle n_2\rangle_{GC}\right ]=0.
\end{equation} 
Therefore the contribution of $\hat S_1$ for $\dot P_1$  cancels out. In general, 
the contribution of $\hat S_1$ for $\dot P_1^n$  is 
not exactly zero, but it is always a constant, following the MD arguments. 
Therefore our system reduces to a two--level 
system, in which only the levels $0$ and $1$ must be considered. 

\subsection{Adiabatic elimination of the level $1$}

Having reduced the system into just two levels, $0$ and $1$, we 
shall adiabatically eliminate the level $1$. This can be achieved because, as 
observed from Eqs. (\ref{P1dot}) and (\ref{P0dot}), $P_1$ decays on a time scale of the order ${\cal O}(1/N)$, 
whereas $P_0$ decays in a time scale of the order ${\cal O}(1)$. We are interested in 
contributions up to the order ${\cal O}(1/N)$ in $\dot P_0$, 
and hence in contributions up to the order ${\cal O}(1/N)$ in the stationary value of $P_1$. 
From Eq. (\ref{P1dot}) it is clear 
that the contributions of the order ${\cal O}(1)$ in $\dot P_1$ 
lead to terms of the order ${\cal O}(1/N)$ in the stationary value of $P_1$. 
Therefore we are interested in $\dot P_1$ up to order ${\cal O}(1)$, and therefore in $P_1^2$ up to order ${\cal O}(1)$. This means that we 
need to calculate $\dot P_1^2$ just up to the order ${\cal O}(N)$:
\begin{eqnarray}
&&\dot P_1^{2}=\frac{\Gamma_-}{2}\left [ -4(1-q)N|x_0|^2 \right ] P_1^{2}  \nonumber \\
&&+\frac{\Gamma_-}{2} \left [ 8 q  N|x_0|^2 \right ]P_1 .
\end{eqnarray}
For times $t\gg [2\Gamma_-(1-q)N|x_0|^2]^{-1}$, the stationary values:
\begin{equation}
P_1^{2}=2\frac{q}{q+1}P_1.
\end{equation}
is obtained. Therefore Eq.\ (\ref{P1dot}) 
reduces to the form (up to order ${\cal O}(1)$):
\begin{eqnarray}
&&\dot P_1=\frac{\Gamma_-}{2}\left [ -2(1-q)N|x_0|^2 +\hat l_0 -2\right ] P_1 \nonumber \\
&&+\frac{\Gamma_-}{2} \left [ 2 q  N|x_0|^2 \right ]P_0.
\end{eqnarray}
And for $t\gg [\Gamma_-(1-q)N|x_0|^2]^{-1}$,
\begin{equation}
P_1\simeq\left [ \frac{q}{1-q}+\frac{q}{2N(1-q)^2}\hat l_o \frac{1}{|x_0|^2} \right ] P_0.
\end{equation}
Substituting this value in Eq. (\ref{P0dot}), one obtains Eq.\ (\ref{P0dot2}).

\begin{references}
\bibitem{Kettcomment} The new $m_F=-1$ Feschbach resonance at 1195 G in Sodium,
  reported in Ref.\ \cite{Ketterle99}, opens the possibility to
  investigate Bose gases with zero scattering length 
  (this value of the scattering length can be obtained without 
  crossing any resonance and therefore avoiding the strongly enhanced
  inelastic collisions). In this situation we would have practically an 
  ideal gas, as considered in the paper.
\bibitem{LDLcomment} Only the transitions $m\rightarrow m+1$ are important. 
  This is because we are in the LDL, and the trap is centered in a node of the 
  laser standing wave.
\end{references}
\end{document}